\documentclass[%
 reprint,
preprintnumbers,
nofootinbib,
 amsmath,amssymb,
 aps,
]{revtex4-2}

\usepackage[dvipdfmx]{graphicx}
\usepackage{amsmath}
\usepackage{amsfonts}
\usepackage{amssymb}
\usepackage{graphicx, rotating}
\usepackage{epsfig}
\usepackage{latexsym}
\usepackage{graphicx}
\usepackage{color}
\usepackage{amsmath,bm,amssymb}

\usepackage{color}
\usepackage[english]{babel}
\usepackage{hyperref}

\usepackage{textcomp}
\usepackage{ulem}

\newcommand{\Slash}[1]{{\ooalign{\hfil#1\hfil\crcr\raise.167ex\hbox{/}}}}
\newcommand{\bra}[1]{ \langle {#1} | }
\newcommand{\ket}[1]{ | {#1} \rangle }
\newcommand{\beq}{\begin{equation}}  \newcommand{\eeq}{\end{equation}}
\newcommand{\bef}{\begin{figure}}  \newcommand{\eef}{\end{figure}}
\newcommand{\bec}{\begin{center}}  \newcommand{\eec}{\end{center}}
  
\newcommand{\laq}[1]{\label{eq:#1}}  

\newcommand{\Eq}[1]{Eq.~(\ref{eq:#1})}

\newcommand{\eq}[1]{(\ref{eq:#1})}

\newcommand{\ab}[1]{\left|{#1}\right|}
\newcommand{\vev}[1]{ \left\langle {#1} \right\rangle }

\newcommand{\SU}[1]{{\rm SU{#1} } }

\def\({\left(}
\def\){\right)}

\def\O{\mathcal{O}}
\def\U{\mathop{\rm U}}

\newcommand{\OR}{~{\rm or}~}
\newcommand{\AND}{~{\rm and}~}

\newcommand{\MEV}{ {\rm \, MeV} }
\newcommand{\GEV}{ {\rm \, GeV} }

\def\a{\alpha}
\def\b{\beta}

\def\d{\delta}
\def\f{\phi}
\def\g{\gamma}
\def\k{\kappa}
\def\l{\lambda}
\def\m{\mu}

\def\G{\Gamma}

\def\F{\Phi}
\def\tl{\tilde}

\begin{document}

\preprint{TU-1149}
\title{
Suppression of Higgs Mixing by Quantum Zeno Effect
 }

\author{Kodai Sakurai}
 \email{kodai.sakurai.e3@tohoku.ac.jp}
\author{Wen Yin}%
 \email{yin.wen.b3@tohoku.ac.jp}
\affiliation{%
Department of Physics, Tohoku University, Sendai, Miyagi 980-8578, Japan
}%

\date{\today}

\begin{abstract}
The Higgs portal interaction to a singlet sector of the standard model (SM) gauge group is widely-studied.
In this Letter, we show that a quantum effect is important if the Higgs field mixes with another singlet scalar field whose decay rate is larger than the mass difference between the two mass eigenstates. This effect may be interpreted as the quantum Zeno effect.  
In either the quantum mechanics or the quantum field theory, we show that the resulting propagating mode is not the eigenstate of the mass matrix, but it is approximately the eigenstate of the interaction. 
As a consequence, the decoupling of the mixing effect happens at the infinity limit of the decay width of the exotic scalar even if the na\"{i}ve mixing parameter is not small. 
With a finite decay width of the exotic scalar, we derive the effective mass of the propagating mode in the SM sector, its decay rate, and the couplings at the 1-loop level. It turns out that the mixed mass eigenstates can mimic the discovered 125~GeV Higgs boson. This fuzzy Higgs boson can be obtained in a simple perturbative renormalizable model. It is consistent with the 125~GeV SM Higgs boson when the mass difference is smaller than $O(0.1)$GeV ($O(1)$GeV) for $O(1)$ ($O(0.01)$) mixing. 
We argue the possible natural scenario for the tiny mass splitting and the possibility that the upper bound of the mass difference is larger for a strongly-coupled singlet sector. 
To probe the fuzzy Higgs boson scenario, it is difficult to directly produce the singlet sector particles.
Nevertheless, the future Higgs factories may probe this scenario by precisely measuring the Higgs boson invisible decay rate and the deviation of the Higgs coupling. Applications of the mechanism are also mentioned.
\end{abstract}

\maketitle


{\bf Introduction.--}
The existence of a gauge singlet sector, which we call the dark sector, is plausible due to the evidence of dark matter. 
In particular, the dark sector may include light particles since the stability is easily guaranteed for light dark matter.
The dark sector may connect with the SM sector via a portal interaction between the Higgs and dark Higgs fields~\cite{Silveira:1985rk,Burgess:2000yq}.

In the broken phase, in general, the Higgs potential is given as,  
$
V=\sum_{i=0}{h^i V_{\rm dark}^{(i)}},$ with $h$ being the Higgs boson that is embedded in the Higgs doublet field. 
$V^{(i)}_{\rm dark}$ is the potential as a function of solely gauge-singlet fields and parameters.  
For concreteness, let us consider the renormalizable potential with $\U(1)$ global and 
$\U(1)_Y\times \SU(2)_L\times \SU(3)_c$ gauge symmetries: 
\beq
V=-m^2_\F|\F|^2+{\lambda} |\F|^4 +\l_P |H|^2 |\F|^2 + \lambda_H |H|^4-\mu_H^2 |H|^2.  \label{V}
\eeq
Here $\F$ ($H$) is the dark (SM) Higgs field, which is gauge singlet (doublet), whose vacuum expectation value (VEV) 
 will break the $\U(1)$ ($\SU(2)_L\times \U(1)_Y$) symmetry; 
$\lambda_P, \l (>0) \AND \l_H (>0)$ are coupling constants;  $\m_H^2\AND m_\F^2$ 
are the bare mass parameters. 
A slight modification of the model has been studied in the context of axion/dark photon dark matter production via the phase transition~\cite{Nakayama:2021avl},
 the UV model of a CP-even ALP~\cite{Sakurai:2021ipp}, 
WIMP~\cite{Barger:2008jx, Barger:2010yn, Gonderinger:2012rd}\cite{Ishiwata:2018sdi, Cline:2019okt, Grzadkowski:2020frj, Abe:2020iph, Abe:2021nih}, electroweak baryogenesis~\cite{Barger:2008jx,Cho:2021itv} and collider physics~\cite{Chen:2019ebq, Grzadkowski:2020frj, Abe:2021nih, Bhattacherjee:2021rml, Sakurai:2021ipp}.

The symmetry breaking occurs because of $\vev{\F}=v_s, \vev{H_0}=v$. 
We obtain a massless Nambu-Goldstone boson (NGB), $a$, from the $\U(1)$ breaking.
This model is not very special for realizing the {mechanism} in this Letter. 
The general and important feature is that we can define the ``flavor", similar to the neutrino oscillation, by the (broken) gauge symmetry, i.e.
$h$ is the SM flavor, and the other, $s$, is the dark flavor:
\begin{align}
H=
\begin{pmatrix}
G^{+} \\
\frac{1}{\sqrt{2}}(v+h+iG^0)
\end{pmatrix},\quad
\Phi=\frac{1}{\sqrt{2}}(v_{s}+s+ia),
\end{align}
where $G^{+}$ and $G^{0}$ are NGBs absorbed by the longitudinal mode of the weak gauge bosons. 
In general, $s$ and $h$ are not mass eigenstates since they mix with each other. 

When the mixing is relatively small, $s$ is almost the mass eigenstate {$\phi_2$} with the mass $m_2$. Thus $s$ can {dominantly} decay into a pair of the {NGBs, $a$}.
The decay rate is $ \Gamma_s[m_2] \equiv  \frac{m^3_2}{32\pi v^2_s}\simeq \frac{\lambda_s m_2}{8\pi}$. 
When $\l_s=\O(1)-4\pi$, {the upper of which is a perturbative unitarity bound},
$\G_s=\O(10\%-1)m_2.$ 
Therefore, there is a parameter region  
\beq \laq{cond} |m_2-m_1| \lesssim \Gamma_s \lesssim m_2, \eeq
when $|m_2-m_1|\lesssim \O(10\%-1) m_2$ is satisfied.
Here $m_1$ is the mass of the other mass eigenstate $\phi_1$. 
The region satisfying \eq{cond} is our focus.
A relativistic state with momentum $p$ has the decay time $t_{\rm s}\sim p/ {(m_2\G_{s})}$
which is shorter than the timescale of $t_{\rm unc}\sim 2p/(|m^2_2-m^2_1|)\sim p/(m_2 |m_2-m_1|)$ for measuring the energy difference according to the uncertainty principle. 
Therefore, the timescale that defines the mass eigenstates as the asymptotic states~e.g. \cite{weinberg1995quantum} does not exist.

To study the system with \eq{cond}, for clarity, we consider the following Higgs production process in an electron-positron collider:
\beq
e^+ e^-\to Zh(=\cos\a \f_1-\sin\a \f_2).
\eeq
We emphasize that $h$ is the  SM flavor eigenstate, which is a superposition of the mass eigenstates, $\f_1 \AND \f_2$.  $\a$ is the mixing angle.

In this Letter, we show that a quantum effect is so important that the on-shell state is {\it NOT} the mass eigenstate $\f_{i}$. This may be interpreted as the quantum Zeno effect e.g. \cite{Nakazato:1995cn} by defining the decay of $s$ as the measurement. It states that the quantum system with frequent measurements evolves in a subspace of the total Hilbert space. 
 The frequent decay of the dark flavor state forces the SM flavor state to evolve within itself, i.e. the Higgs mixing interactions with the dark Higgs boson are decoupled. 
 To distinguish from the conventional SM Higgs boson, we call the propagating SM flavor state, fuzzy Higgs boson.
We derive the property of the fuzzy Higgs boson: the mass, the decay rate, and the couplings, and show that it can be consistent with the SM Higgs boson, and thus the discovered Higgs boson {with the mass of 125 GeV} might be the fuzzy one. 
It is key to probe this scenario by precisely measuring the Higgs boson property.

Let us list some relevant studies. 
In the context of the physics beyond the SM (BSM), the quantum effect leading to the invalid use of na\"{i}ve asymptotic state was studied widely (Within the SM plus neutrino masses, the neutrino oscillation and meson oscillation phenomena belong to this category.)
For instance, there are baryogenesis mechanisms via right-handed neutrino oscillations~\cite{Akhmedov:1998qx,Asaka:2005pn, Canetti:2012kh} 
and via the left-handed neutrino oscillations~\cite{Hamada:2016oft, Hamada:2018epb,Eijima:2019hey} (see also quark flavor oscillations~\cite{Asaka:2019ocw} and hadron oscillatoins~\cite{McKeen:2015cuz, Aitken:2017wie, Elor:2018twp}).
Oscillation effects of axions~\cite{Chadha-Day:2021uyt} or gauge bosons~\cite{Cacciapaglia:2009ic} motivated by extra-dimensional models are also interesting.  
{In \cite{Arkani-Hamed:1997pqv, Cacciapaglia:2009ic}}, it was pointed out that appropriate treatment of the Breit-Wigner formulation~\cite{Pilaftsis:1997dr} is necessary in case the mass difference between two heavy scalar (vector) bosons is close to their widths.
In \cite{Arkani-Hamed:1997pqv,Fuchs:2014ola,Fuchs:2016swt,Bian:2017jpt,Das:2017tob,Das:2018haz,Das:2020ujo,MoosaviNejad:2019agw}, {the phenomenological implications of the interference effects of multiple nearly-mass degenerate scalar particles with same gauge charges, e.g. Higgs doublets and sleptons, are studied.}
In \cite{Grzadkowski:2020frj}, {applying the Breit-Wigner formulation, a similar model to our setup} was investigated with MeV decay width of $s$ but the region of our interest was not the authors' focus. 
{What is new in our setup is that the dark flavor state has a large decay width. As far as we know, the suppression of the Higgs mixing effect by the large width has not been clearly pointed out and the resulting fuzzy Higgs property has not been well studied. }

It was noticed that 
the nearly degenerate two Higgs bosons are favored to have consistent cosmology and low-energy phenomenology~\cite{Abe:2021nih, Cho:2021itv}. 
{In particular, the current LHC does not have enough detector resolution to constrain the decays of the degenerate Higgs bosons into the SM particles whose widths together are similar to that of the SM Higgs boson~\cite{CMS:2014afl}.
In contrast, by omitting the quantum Zeno effect, this is not true in our setup, because the two mass eigenstate bosons would decay invisibly due to the mixing. Thus both the SM interaction and the invisible decay branching fraction would be constrained. 
We will discuss that the constraints can be alleviated by increasing the decay width of the dark flavor scalar. }

{\bf Intuitive discussion from quantum mechanics.--}
Let us first discuss the system in an intuitive way by using quantum mechanics to relate the quantum Zeno effect. The reader, who would like to have rigid discussions, should move to the next part.

We define the SM flavor state as 
 $\ket{h}$, which only couples to the SM particles, and the dark flavor state as $\ket{s}$, who only couples to the dark sector particles at the vanishing limit of $\lambda_P$.
The mass eigenstates are defined as $\ket{i}$ with $i=1,2$. 
Then the mixing parameter is defined to satisfy
$
\cos \a\equiv \bra{1}h\rangle, \sin\a \equiv \bra{2}h \rangle, \bra{1}s\rangle=-\sin \a, \bra{2} s\rangle=\cos \a.
$

Now we can follow the time evolution of $\ket{h}$ with the relativistic momentum  $\vec p$ at the time $t=0$. 
We obtain the probability to measure  $\ket{s}$ flavor state at the time $t$ as
\begin{align}
P[t]&=\ab{\bra{s,\vec p} \exp{[-i \hat{H}_{\rm free} t]} \ket{h,\vec p}}^2\notag \\ 
&\approx 4\cos^2 \a \sin^2 \a \sin^2{\(\frac{\epsilon}{4|\vec p|} t\)} \notag\\
& {\rm with~} \epsilon  \equiv m_1^2-m_2^2
\end{align}
where $\hat{H}_{\rm free}$ is the free Hamiltonian, which defines the mass eigenstates of $\ket{1} \AND \ket{2}$. The eigenvalues 
are $\sqrt{\vec p^2+ m_1^2}, \AND \sqrt{ \vec p^2+m_2^2} $, respectively. In the last equality, we have expanded $m_i$. 
We note that the quantum mechanics treatment is only valid at a short enough timescale. 
When $t$ is longer than the life-time of  $s$, $t_{s}\sim |\vec p|/ (\bar{m}_s\G_s[\bar{m}_s])$, 
we have to take account of the $s$ decay.  Here $\bar{m}_s$ is the effective mass of the dark flavor and it is a certain average of $m_i$. 
If we define the decay as the measurement of $\ket{s}$,  we may consider that the measurements take place in the period of $\sim t_s$. The measurement $\ket{s}$ with the probability of 
\beq
P[t_s] \approx 4 \cos^2 \a \sin^2 \a \(\frac{\epsilon}{4\bar{m}_s \G_s[\bar m_s]}\)^2.
\eeq
Here we have used that $t_s \lesssim |\vec p|/\ab{\epsilon}.$ $h$ also decays into the SM particles in reality, but we neglect it for illustrative purposes
(This is included in the following numerical estimation.)

At $t=t_s$, the SM flavor is measured with the probability $1-P[t_s]$.
Later, the quantum state starts to evolve again from $\ket{h}$. 
In every $t_s$ step this measurement happens. 
Therefore at $t\gg t_s$ we can measure the SM flavor with the probability 
\beq\laq{ph}
P_h[ t(\gg t_s)]\sim (1-P[t_s])^{\frac{t}{t_s}} \sim \exp{[-P[t_s] \frac{t}{t_s}]}.
\eeq
If $t\ll t_s/P[t_s]$, i.e. the measurement of $\ket{s}$ is too frequent, one finds $\ket{h}$ at any time of its propagation. 
This can be interpreted as the quantum Zeno effect 
due to the multiple frequent ``measurements" via the $\ket{s}$ decay.

The energy of $\ket{h}$ is obtained as 
$\bar E_h\simeq \bra{h,\vec p} \hat{H}_{\rm free} \ket{h,\vec p}
\simeq |\vec p|+ \frac{m_1^2\cos^2\a +{m_2^2\sin^2\a} }{2|\vec p|}.
$ 
 Thus we get the effective mass of the propagating state, $m^2_{h\rm eff}\equiv \bar E_h^2 -\vec p^2$, as
\beq\laq{mass}
\boxed{m_{h\rm eff} \simeq  \sqrt{m_1^2 \cos^2\a+m_2^2\sin^2 \a}}. 
\eeq 

From \Eq{ph} we find that the SM flavor eigenstate has a decay rate of 
\beq 
\laq{width}
\Gamma_{h\to {\rm dark}}\sim \frac{p}{m_{h\rm eff}} \frac{P[t_s]}{t_s}.
\eeq
On the contrary, 
in the conventional perturbation theory, the invisible decay rate of the Higgs flavor is estimated as 
$  \frac{\(|\bra{s}1\rangle \langle 1 \ket{h}|^2+ |\bra{s}2\rangle \langle 2\ket{h}|^2\)}{t_s}\sim  2\cos \a^2 \sin\a^2/t_s $ with $m_1\sim m_2.$ 
Our result gives a further suppression by $\sim  \(\frac{\epsilon}{4\bar{m}_s \G_s[\bar m_s]}\)^2$.  
In the following, we will confirm this intuitive discussion by quantum field theory.

{\bf Breit-Wigner formulation.--} 
To study the propagating mode, we can estimate the full two-point function. 
To this end, we estimate the resummed propagator. 
The propagator in the flavor basis is 
\beq
{\hat{\Delta}= i R\cdot \begin{pmatrix}
Q^2-m_1^2+\hat{\Pi}_{11} & \hat{\Pi}_{12} \\
\hat{\Pi}_{21} &  Q^2-m_2^2+\hat{\Pi}_{22}\\
\end{pmatrix}^{-1} \cdot R^T.}
\eeq
Here $\hat{\Pi}$ is the renormalized self-energy in the mass basis, and 
$R_{h1}\equiv\bra{h}1\rangle=\cos\a$,\;$R_{h2}\equiv\bra{h}2\rangle=-\sin\a$ and so on. 
We perform the 1-loop calculation of the self-energy to obtain
\beq
\hat{\Pi}_{ij}\simeq \frac{i}{32\pi v_s^2} \begin{pmatrix}
\sin^2(\a) m_1^4 & \sin\a \cos\a m_1^2 m_2^2  \\ 
\sin\a \cos\a m_1^2 m_2^2  & \cos^2(\a) m_2^4 \\
\end{pmatrix}
\eeq
where we used the on-shell renormalization (c.f.~\cite{Krause:2016oke,Kanemura:2017wtm}), and thus the real part is dropped by the renormalization conditions.  
Here we omit to write down the subdominant contribution of the imaginary part from the Higgs decay into the SM particles for clarity. 
The full 1-loop formulas are given in Appendix A.

According to the LSZ reduction or optical theorem, the mass of the particle corresponds to the real part of the pole of the resummed propagator, 
 and the decay rate of the propagating mode can be obtained from the imaginary part. 
Since we are interested in $|\epsilon|= |m^2_1-m^2_2|  \ll \Gamma_s m_{h\rm eff}\sim \frac{m_{h\rm eff}^4}{32\pi v_s^2 }$, we expand the denominator, $\det [(-i \hat{\Delta})]^{-1}$, of the propagator with $\epsilon$.  
By noting $m_1^2=m_{h\rm eff}^2 + \epsilon\cos^2\a , m_2^2=m_{h\rm eff}^2 - \epsilon\sin^2\a,$ 
the denominator factorizes as
\begin{align}
\text{denominator of } \hat{\Delta} \simeq 
 \laq{1}
(Q^2-m_{h \rm eff}^2 +\epsilon \cos(2\alpha)+i\frac{m_{h \rm eff}^4}{32\pi v_s^2})\\
\times (Q^2-m_{h \rm eff}^2+\frac{\epsilon^2 \sin^2(2\alpha)}{2m_{h\rm eff}^2}+i \frac{8 \pi  \epsilon^2 \sin^2(2\alpha ) v_s^2}{  m_{h\rm eff}^4}). \laq{2}
\end{align}
 Here we have neglected higher-order terms in $\epsilon$ that are irrelevant for analytic estimation. 
 This is common 
 in any component of $\hat{\Delta}_{\b\g}~(\b,\g=h,s).$ 
The real parts of the two poles are separated by $\O(\epsilon)$, which is smaller than the imaginary part of \Eq{1}, $\frac{m_{h \rm eff}^4}{32\pi v_S^2}\simeq \G_s {m}_{h\rm eff}$. 
We can also find 
\beq
\text{numerator of } \hat{\Delta}_{hh}\simeq i(Q^2-m_{h \rm eff}^2 +\epsilon \cos(2\alpha)+i\frac{m_{h \rm eff}^4}{32\pi v_s^2})
\eeq
which is the same as \eq{1} at the leading order of $\epsilon$ in either real and imaginary part. 
As a consequence, we obtain
\beq\laq{prop}
\hat\Delta_{hh} \simeq \frac{i(1+\delta)}{Q^2-m_{h \rm eff}^2+\frac{\epsilon^2 \sin^2(2\alpha)}{2m_{h\rm eff}^2}+i \frac{8 \pi  \epsilon^2 \sin^2(2\alpha ) v_s^2}{  m^4_{h\rm eff}}}.
\eeq
Here $\d$ denotes the higher order correction of order $\epsilon^2$. For instance, we obtain
 $ \delta(Q^2 \simeq m^2_{h\rm eff})\simeq-
 \sin ^2(2 \alpha ) \frac{\epsilon ^2\left({m_{h \rm eff}^2+32 i \pi  v_s^2}\right)^2}{4 m_{h\rm eff}^8}$
 around the pole while it is further suppressed by $Q^2-m^2_{h\rm eff}$ far away from the pole. 
 
Interestingly $\hat{\Delta}_{hh}$ behaves as a propagator with a single pole.
From the real part, we obtain that the mass of the propagating mode to be
 {$m_{ h \rm eff}^2$} with a tiny correction of $\O(\epsilon^2/m^2_{h \rm eff})$ consistent with \Eq{mass}. 
 The decay rate into the dark sector is read from the imaginary part,
\beq \boxed{\Gamma_{h{\rm eff}\to \rm dark} \simeq \frac{\epsilon^2 \sin(2\a)^2}{4m^2_{h\rm eff}\Gamma_s[m_{h\rm eff}]}}\eeq which is consistent with \Eq{width}. 
${|\hat{\Delta}_{sh, hs}(Q^2\sim m^2_{h{\rm eff}})|}\simeq  \ab{\frac{\epsilon\sin(2\a)}{2{m_{h \rm eff}}\Gamma_{s}[ m_{h\rm eff}]} \frac{i}{Q^2-m^2_{h\rm eff}}}$ around the pole.  
$\hat{\Delta}_{ss}$ includes the two poles.
The numerator of $\hat\Delta_{hh} \AND \hat\Delta_{hs}$ at around the pole is similar to the system that the Higgs mixes with a (decoupled) heavy particle decaying into the dark particles c.f. \cite{Sakurai:2021ipp} by defining the effective mixing angle
\beq \boxed{\sin{[2\alpha_{\rm eff}]}\simeq \frac{\epsilon \sin(2\a)}{m_{h\rm eff}{\Gamma_{s}[m_{h\rm eff}]}}},\eeq  
by noting that the numerator of $\hat\Delta_{hs}$ ($\hat\Delta_{hh}$) around the pole is $\sin\alpha_{\rm eff}\cos\alpha_{\rm eff}$ ($ \cos^2[\alpha_{\rm eff}]$) at the leading order.
This expresses that the  fuzzy Higgs boson coupling to the SM particles is 
$g_{h_{\rm eff}XX}^{\rm SM}= g^{\rm SM}_{hXX}\cos[\alpha_{\rm eff}]$,  Here $g^{\rm SM}_{hXX}$ is the tree-level Higgs boson coupling within the SM.

\begin{figure}[!t]
   \includegraphics[width=.43\textwidth]{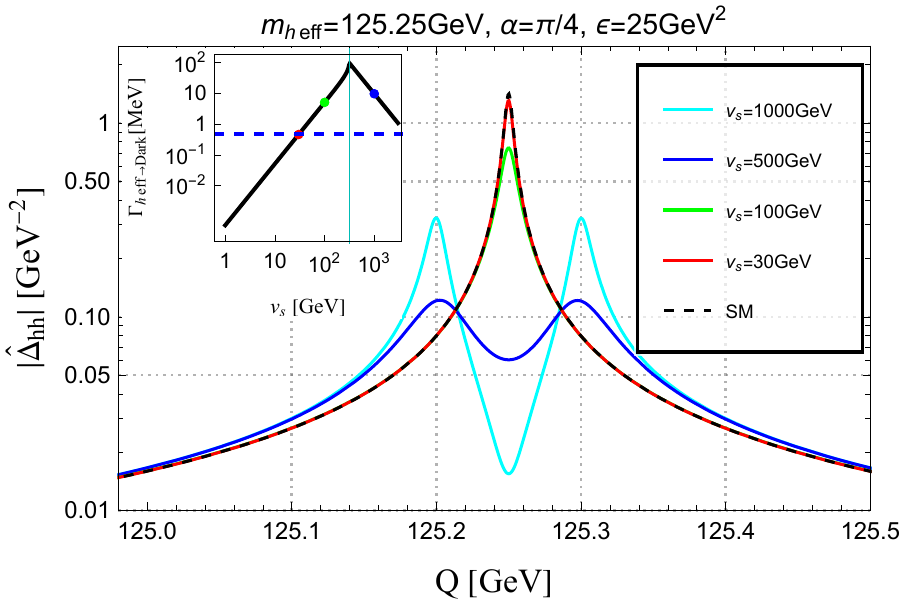}
   \includegraphics[width=.43\textwidth]{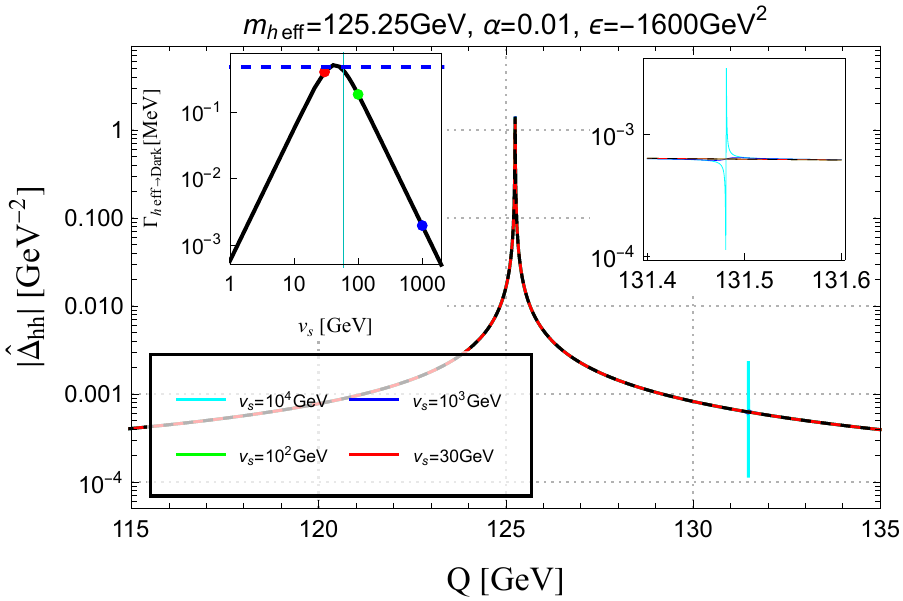}
\caption{The {(h,h)} element of the resumed propagator, $|\hat\Delta_{hh}|$, by varying $Q^2$.  
In the top (bottom) panel, we fix $\alpha=\pi/4~({0.01}),~m_{h\rm eff}=125.25\GEV,~{\epsilon=25\GEV^2}~({-1600\GEV^2})$ and plot the case $v_s=10^3,\;500,\;100,\;{30}\GEV~(10^4,\;10^3,\;100,\;30\GEV)$ in cyan, blue, green, red solid lines, respectively. 
These inputs yield { $m_1=125.3\GEV$ {($125.25\GEV$)} and $m_2=125.2\GEV$} ({$131.5\GEV$}) in the top (bottom) panel.
Also shown is the resumed pure SM Higgs propagator in the black dashed line. 
{The corresponding decay widths of the fuzzy Higgs boson are shown in built-in figures with the horizontal blue dashed line being the current bound. The colored points correspond to the same colored lines for $|\hat\Delta_{hh}|.$
The vertical cyan line in each figure denotes $\Gamma_s[m_{h\rm eff}]\approx |m_1-m_2|$ for comparison.}} \label{fig:0} 
\end{figure}

{\bf Fuzzy Higgs boson as the 125\,GeV one.--}
 Including full 1-loop corrections, we plot $|\hat\Delta_{hh}|$ in Fig.\ref{fig:0} by varying $Q$ for $\alpha=\pi/4, \epsilon =25\GEV^2$ and {$\alpha=0.01, \epsilon =-1600\GEV^2$} in the top and bottom panels, respectively, in colored solid lines. 
In the top (bottom) figure, we fix $ m_{h\rm eff}=125.25$\,GeV with
 $v_s=1000, 500,  100, \AND {30} \GEV$ ($10^4, 10^3, 100, \AND 30\GEV$). Decreasing $v_s$ corresponds to the increase of $\G_s$. One can see that the two peaks at large $v_s$ become a single one by decreasing $v_s$, while the single peak position approaches to $m_{h\rm eff}.$ 
We also denote the usual SM Higgs propagator by the black dashed line. 
When $v_s$ gets smaller, $\hat{\Delta}_{hh}$ approaches to the SM Higgs one.

Also shown in the built-in panels are the peculiar features of the fuzzy Higgs decay width into $aa$. 
This is estimated from the width of the propagator involving only $a$ loops. The colored points correspond to the same colored lines for $|\hat\Delta_{hh}|.$
The horizontal blue-dashed line corresponds to the current constraint of the branching fraction by ATLAS, 11\%~\cite{ATLAS-CONF-2020-052} (see also Refs.~\cite{ATLAS:2019cid, CMS:2018yfx}). Thus the invisible decay can be consistent with the observation with sufficiently small $v_s$. 

 In the region satisfying the invisible decay bound in the two figures, the size of effective mixing $\sin[\alpha_{\rm eff}]^2\cos[\alpha_{\rm eff}]^2\simeq \Gamma_{h\rm eff\to dark}/\G_s[m_{ h\rm eff}]\lesssim {\rm \MEV}/\G_s[m_{ h\rm eff}]$ is tiny and 
 the deviations of the couplings are very small~(c.f. \cite{Aad:2019mbh, CMS:2020gsy}).
There are also other loop corrections, rather than the imaginary part contribution, to the fuzzy Higgs coupling but they are even smaller (see Appendix B). Thus the discovered Higgs boson can be consistent with the fuzzy Higgs boson who is the superposition of the mass eigenstates. 

Since it is difficult to produce a resonance absent in $\hat{\Delta}_{hs} \OR \hat{\Delta}_{hh}$ in a SM particle collider, detecting the light $s$ may not be easy.\footnote{That said, $s$ may be produced via the $h_{\rm eff}ss$ or $h_{\rm eff}h_{\rm eff}ss$ vertex from the portal coupling. But this cross-section is small due to the small $|\lambda_P|$ (Note that we have either small $\epsilon$ or $\alpha$.)  } 
One way to probe this scenario may be the Higgs invisible decay. 
Therefore probing this scenario is favored in the future lepton colliders such as FCC-ee, CEPC, ILC, CLIC~\cite{Asner:2013psa,dEnterria:2016sca,Abramowicz:2016zbo,CEPC-SPPCStudyGroup:2015csa}.\footnote{
Giving mass to $a$ may induce the $a$ decay into the SM particles, which give interesting other signatures~\cite{Sakurai:2021ipp}. } 
When the mass difference is smaller or around $\MEV$, $\G_s$ does not need to be much larger than $\O(\MEV)$. In this case, the effective mixing can get larger, and the fuzzy Higgs boson can be also probed by precisely measuring its couplings.

So far, for concreteness,  the mechanism is explained by using a simple elementary particle model that is perturbative.
The perturbative unitarity gives a lower limit $v_s \gtrsim 25\GEV$ with the criterion $|a_0^i|<1$ for the eigenvalues $a_0^i$ of scattering matrix for the scalar bosons~\cite{Hektor:2019ote}. Since the red points corresponding to $v_s=30\GEV$ are close to the experimental bounds in Fig.~\ref{fig:0}, 
 the fuzzy Higgs parameter region has the mass difference smaller than  $|\epsilon|/m_{h \rm eff}=\O(0.1)\GEV, \AND \O(1)\GEV$ for $\alpha=\O(1) \AND \O(0.01)$ respectively.

{\bf Application to other dark sectors.--} An example of a natural perturbative model for the fuzzy Higgs boson can be obtained with a mirror SM, i.e.  
a dark sector that the SM transforms into via an exact $Z_2$ mirror symmetry. 
In this case, the dark Higgs decay width is around $\sim 4\MEV$ and the mixing is $\alpha=\pi/4$ by introducing a small portal coupling, $\l_P$. The splitting of the two Higgs boson eigenmasses gets smaller with smaller $|\l_P|$. To evade the invisible decay bound, $\sin(\alpha_{\rm eff})\lesssim 0.2$. 
Thus, the mirror SM can be probed via the invisible decay and the coupling measurement of the fuzzy Higgs boson.

We expect, on the other hand, that the upper bound of the mass difference can be relaxed if $s$ connects a strongly-coupled dark sector e.g. by considering that dark sector particles are composite. Then the perturbative unitarity is not needed. 
Even if the dark sector is strongly coupled, the SM  perturbativity should be under control since all the radiative contributions to the SM sector are suppressed if the portal coupling is small (see c.f. Appendix B).

 {\bf  Applications to cosmology.--} 
 In various cosmological scenarios,
  like baryogenesis, mediator, and light dark matter, one may have interesting parameter regions with the quantum Zeno effect in the vacuum.
In the early Universe, however,
the quantum Zeno effect may be ineffective due to the larger thermal/effective mass differences.

The discussion in this Letter can be easily extended to other models with multiple particles.
For instance, there are models describing the early Universe cosmology with many axions with one of them having a large decay rate/dissipation rate for reheating e.g.~\cite{Daido:2017wwb, Carta:2020oci}. In this case, the quantum Zeno effect may be important. Alternatively, a similar effect may explain the stability of the dark matter. For instance, we can consider two mass eigenstate axions, $a_1,\AND a_2$, both decaying into photons via the couplings $(\cos\b a_1+ \sin \b a_2) g_{a\g\g} F \tl F$ with  $F$ ($\tl F$, $g_{a\g\g} $ ) being the photon field strength  (its dual, photon coupling) and $\beta$ being the mixing angle. If the mass difference is much smaller than the decay rate, then $A\equiv (-\sin \b a_1+\cos \b a_2)$, which is not the mass eigenstate of the Lagrangian parameter, behaves as the propagating mode. In addition, it is stabilized in the case the mass is close. Thus $A$ can become long-lived dark matter. Axion mass difference may depend on temperature such that the axions both decay in the early Universe. This time-dependent decay effect may provide a production mechanism of the axion abundance similar to that in the ALP miracle scenarios~\cite{Daido:2017wwb,Daido:2017tbr} (see also \cite{IAXO:2019mpb, Choi:2019osi, Takahashi:2019qmh, Takahashi:2020uio, Takahashi:2021tff}). 

{\bf Conclusions.--} The Higgs portal interaction to a dark sector of the standard model gauge group is widely-studied. In this Letter, we have shown that a quantum effect is important if the Higgs field mixes with another singlet scalar field whose decay rate is larger than the mass difference between the two mass eigenstates. The resulting propagating mode is approximately the eigenstate of the interaction, and it is different from the eigenstates of the mass matrix. 
As a consequence, the mixing interaction is suppressed with a sufficiently large decay width of the exotic scalar even if the na\"{i}ve mixing parameter is not small. 
From the derived property, it turned out that the mixed mass eigenstates can mimic the discovered 125~GeV Higgs boson, giving a possibility that it is a single fuzzy Higgs boson.
The future Higgs factories are important to reveal whether the 125\,GeV Higgs boson is a fuzzy one by precisely measuring the Higgs boson invisible decay rate and the deviation of the Higgs coupling.  The mechanism also has various applications.
\\
\\

{\bf Acknowledgement.--}
We thank Gi-Chol Cho, Yukinari Sumino and Masahiro Hotta for useful discussions.
This work was supported by JSPS KAKENHI Grant Nos.  20H01894 (K.S.) 20H05851 (W.Y.), 21K20363 (K.S.), 21K20364 (W.Y.), 22K14029 (W.Y.), and 22H01215 (W.Y.).

\clearpage
\appendix

\section{Analytical formulae of the self-energy for Higgs bosons }
The analytical expressions for the self energies $\Pi_{ij}$ $(i,j=1,2)$ with full 1-loop corrections in the mass basis are presented.
In the calculation of $\Pi_{ij}$, We choose the ’t Hooft-Feynman gauge. 
They are given in terms of the Passarino–Veltman functions~\cite{Passarino:1978jh}. 
We first separate $\Pi_{ij}$ to each loop contribution as
\begin{align}
\Pi_{ij}=\sum_f\Pi_{ij}^f
+\Pi_{ij}^{V}+\Pi_{ij}^{\rm PT}+\Pi_{ij}^{S}+\Pi_{ij}^{a}.
\end{align}
Each ingredient is given by
\begin{align}
(16\pi^2)\Pi_{ij}^f &= 
-R_{hi} R_{hj} \frac{4N_f^cm_f^2}{v^2} \Big\{A(m_f)  \notag \\
&+\left(2 m_f^2-\frac{Q^2}{2}\right) B_0(Q^2,m_f,m_f)\Big\}, \\
(16\pi^2)\Pi_{ij}^V&=
\frac{1}{2} g^2R_{hi} R_{hj}  \left(\frac{8}{\delta _{ij}+1}-1\right)A(m_W) \notag\\
&+\frac{1}{4} g_Z^2R_{hi} R_{hj}  \left(\frac{8}{\delta _{ij}+1}-1\right)A(m_Z) \notag \\ 
&-\frac{1}{2} g^2 R_{hi} R_{hj} \left(\frac{1}{\delta _{ij}+1}+1\right)4 m_W^2 \notag\\
&-\frac{1}{4} g_Z^2 R_{hi} R_{hj} \left(\frac{1}{\delta _{ij}+1}+1\right)4 m_Z^2 \notag\\
&-(\delta _{ij}+1)\lambda_{ijG^+G^-}A(m_W)  \notag\\
&-(\delta _{ij}+1)\lambda_{ijG^0G^0}A(m_Z) \notag\\
&+g^2R_{hi} R_{hj} (3 m_W^2-Q^2) B_0(Q^2,m_W,m_W) \notag\\
&+\frac{1}{2} g_Z^2 R_{hi} R_{hj} (3 m_Z^2-Q^2) B_0(Q^2,m_Z,m_Z) \notag\\ 
&+\lambda_{iG^+G^-} \lambda_{jG^+G^-} B_0(Q^2,m_W,m_W) \notag\\
&+2 \lambda_{iG^0G^0} \lambda_{jG^0G^0} B_0(Q^2,m_Z,m_Z),\\
(16\pi^2)\Pi_{ij}^S&=
-(c_1)_{ij}\lambda_{ij11} A(m_1) 
-(c_2)_{ij}\lambda_{ij22} A(m_2)  \notag \\ 
&+(c_{11})_{ij}\lambda_{i11} \lambda_{j11} B_0(Q^2,m_1,m_1)  \notag \\ 
&+(c_{22})_{ij} \lambda_{i22} \lambda_{j22} B_0(Q^2,m_2,m_2) \notag \\ 
&+4 \lambda_{i12} \lambda_{j12} B_0(Q^2,m_1,m_2), \\
(16\pi^2)\Pi_{ij}^{a} &= 
2 \lambda_{iaa} \lambda_{jaa} {B_0(Q^2,0,0)} \notag \\
&-(\delta_{ij}+1)\lambda_{ijaa}{A(0)},
\end{align}
 where $N_f^c$ is the color factor for the fermions, $N_f^c= 3~(1)$ for quarks (leptons) and {the orthogonal matrix $R$ is defined in the main text}. 
The coefficients for $c_m$, $c_{mn}$ $(m,n=1,2)$ are defined by 
\begin{align}
&c_1=
\begin{pmatrix}
12 & 3\\
3 & 2\\
\end{pmatrix}, \quad
c_2=
\begin{pmatrix}
2 & 3\\
3 & 12\\
\end{pmatrix}, \\
&c_{11}=
\begin{pmatrix}
18 & 6\\
6 & 2\\
\end{pmatrix}, \quad
c_{22}=
\begin{pmatrix}
2 & 6\\
6 & 18\\
\end{pmatrix}.
\end{align}
The pinch term contributions $\Pi_{ij}^{\rm PT}$ are given by~\cite{Kanemura:2017wtm,Krause:2016oke}
\begin{align}
(16\pi^2)\Pi^{\rm PT}_{ij}&=
-\frac{1}{4} R_{hi} R_{hj} (2 Q^2-m_i^2-m_j^2) \notag \\
&\times\left\{2 g^2 B_0(Q^2,m_W,m_W)+g_Z^2 B_0(Q^2,m_Z,m_Z)\right\}. 
\end{align}

We also give analytical formulae for the 1-point function for $\phi_1$ and $\phi_2$, which is schematically given by 
\begin{align}
T_i=\sum_{f}T_i^f+T_i^V+T_i^S+T_i^{a}. 
\end{align}
Each contribution is presented by
\begin{align}
(16\pi^2)T^f_i&=
-R_{hi}\frac{4N_f^c m_f^2}{v} A(m_f), \\
(16\pi^2)T^V_i&= \notag \\
&+3R_{hi} g m_W A(m_W) +\frac{3}{2}R_{hi} g_Z m_Z A(m_Z) \notag \\
&-2R_{hi} \left(g m_W^3+\frac{g_Z m_Z^3}{2}\right) \notag \\
&-\lambda_{iG^+G^-}A(m_W) -\lambda_{iG^0G^0}A(m_Z),  \\
(16\pi^2)T^{\rm S}_i&=-\sum_{k=1}^2(b_k)_{i} \lambda_{ikk} A(m_k),  \\
(16\pi^2)T^{a}_i&=-\lambda_{i{\rm aa}}{A(0)}, 
\end{align}
where the coefficient $b_m$ $(m=1,2)$ satisfies
\begin{align}
    b_1=(3,1),\quad b_2=(1,3). 
\end{align}
The scalar couplings are given by
\begin{align}
\lambda_{1G^0G^0}&=
-\frac{m_1^2 \cos \alpha}{2 v}, \\
\lambda_{2G^0G^0}&=
\frac{m_2^2 \sin \alpha}{2 v}, \\
\lambda_{iG^+G^-}&=
2\lambda_{iG^0G^0},\\
\lambda_{111}&=
-\frac{m_1^2 }{2 v v_s}(v \sin ^3\alpha+v_s \cos ^3\alpha), \\
\lambda_{112}&=
\frac{ (2 m_1^2+m_2^2) }{4 v v_s}\sin (2 \alpha )(v_s \cos \alpha-v \sin \alpha), \\
\lambda_{122}&=
-\frac{ (m_1^2+2 m_2^2) }{4 v v_s}\sin (2 \alpha )(v \cos \alpha+v_s \sin \alpha), \\
\lambda_{222}&=
-\frac{m_2^2 }{2 v v_s}(v \cos ^3\alpha-v_s \sin ^3\alpha), \\
\lambda_{1aa}&=
-\frac{m_1^2 \sin \alpha}{2 v_s}
, \\
\lambda_{2aa}&=
-\frac{m_2^2 \cos \alpha}{2 v_s}
,
\end{align}
\begin{align}
\lambda_{11G^0G^0}=&
-\frac{\cos \alpha}{4v^2v_s}\Big[
 (m_1^2-m_2^2)v \sin ^3\alpha \notag\\
&+m_1^2 v_s \cos ^3\alpha +m_2^2 v_s \sin ^2\alpha \cos \alpha \Big] \\
\lambda_{12G^0G^0}=&
\frac{\sin \alpha \cos \alpha}{2v^2v_s}
\Big[
 (m_2^2-m_1^2)v \sin \alpha \cos \alpha \notag\\
 &+m_1^2 v_s \cos ^2\alpha+m_2^2 v_s \sin ^2\alpha
\Big] \\
\lambda_{22G^0G^0}=&
-\frac{\sin \alpha}{4v^2v_s}\Big[
 (m_1^2-m_2^2)v \cos ^3\alpha  \notag \\
 &+m_1^2 v_s \sin \alpha \cos ^2\alpha +m_2^2 v_s \sin ^3\alpha
\Big], \\
\lambda_{ijG^+G^-}&=2\lambda_{ijG^0G^0}, 
\end{align}
\begin{align}
\lambda_{1111}=&
\frac{1}{8v^2v_s^2}\Big[
-m_1^2 v^2 \sin ^6\alpha-m_1^2 v_s^2 \cos ^6\alpha \notag \\
&-m_2^2 v^2 \sin ^4\alpha \cos ^2\alpha-m_2^2 v_s^2 \sin ^2\alpha \cos ^4\alpha
\notag \\
&-2(m_1^2-m_2^2) v v_s \sin ^3\alpha \cos ^3\alpha
\Big],  \\ 
\lambda_{1211}=&
-\sin (2 \alpha ) \frac{v \sin \alpha-v_s \cos \alpha}{16v^{2}v_{s}^{2}}\Big[   \notag\\
&v  (3 m_1^2+m_2^2)\sin \alpha \notag\\
&+v (m_2^2-m_1^2)\sin (3 \alpha ) \notag\\
&+v_s  (3 m_1^2+m_2^2)\cos \alpha\notag\\
&+v_s  (m_1^{2}-m_2^{2})\cos (3 \alpha ) \Big], \\
\lambda_{2211}=&
-\frac{\sin \alpha \cos \alpha}{32v^{2}v_{s}^{2}} \Big[ \notag \\
&6 \sin (2 \alpha ) (m_1^2+m_2^2) (v^2+v_s^2) \notag \\
&-(m_1^{2}-m_2^{2})  (3  (v^{2}-v_s^{2})\sin (4 \alpha ) -2 v v_s) \notag \\
&+6 v v_s  (m_1^{2}-m_2^{2})\cos (4 \alpha )  \Big],
\end{align}
\begin{align}
\lambda_{1122}&=\lambda_{2211},\\
\lambda_{1222}&=\lambda_{1211} \notag \\
+&\frac{\sin (4 \alpha )}{16v^2v_s^2} \Big[ \cos (2 \alpha ) (m_1^2-m_2^2) (v^2-v_s^2) \notag \\
&-(m_1^2+m_2^2) (v^2+v_s^2)\notag \\
&+2 v v_s (m_1^2-m_2^2)\sin (2 \alpha ) \Big]
\\
\lambda_{2222}&=\lambda_{1111}\notag \\
&-\frac{\cos (2 \alpha )}{8v^{2}v_{s}^{2}} \Big[\sin ^2\alpha (m_1^2 v^2-m_2^2 v_s^2) \notag\\
&+\cos^2 \alpha (m_2^2 v^2-m_1^2 v_s^2) \Big],
\end{align}
\begin{align}
\lambda_{11aa}&=
-\frac{\sin \alpha} {4v v_{s}^{2}}
\Big[
v_s \cos ^3\alpha (m_1^2-m_2^2) \notag \\
&+v \sin \alpha (\sin ^2\alpha (m_1^2-m_2^2)+m_2^2)
\Big], \\
\lambda_{12aa}&=
-\frac{\sin \alpha\cos \alpha}{2v v_{s}^{2}}  \Big[
 (m_2^{2}-m_1^{2})v_s \sin \alpha \cos \alpha  \notag \\
&+m_1^2 v \sin ^2\alpha+m_2^2 v \cos ^2\alpha
\Big],\\
\lambda_{22aa}&=
-\frac{\cos \alpha }{4v v_s^2}
\Big[v_s  (m_1^2-m_2^2)\sin^3\alpha\notag \\
&+m_1^2 v \sin ^2\alpha \cos \alpha +m_2^2 v \cos ^3\alpha
\Big].
\end{align}
\section{One-loop contributions to the deviations in the Higgs boson coupling}
We here evaluate the one-loop corrections to the coupling constants for the fuzzy Higgs boson $h_{\rm eff}$ with any SM particle $X$, considering the case of $|m_{2}^{2}-m_{1}^{2}|\ll m_{h {\rm eff}}\Gamma_{s}$. 
The renormalized vertex function can be schematically expressed by 
\begin{align}
\Gamma_{h {\rm eff}XX}=\Gamma^{\rm tree}_{h {\rm eff}XX}+\Gamma^{\rm 1PI}_{h {\rm eff}XX}+\delta\Gamma^{}_{h {\rm eff}XX},
\end{align}
where $\Gamma^{\rm tree}_{h {\rm eff}XX}$ denotes the tree-level coupling $\Gamma^{\rm tree}_{h {\rm eff}XX}$  and  $\Gamma^{\rm 1PI}_{h {\rm eff}XX}$ ($\delta \Gamma^{}_{h {\rm eff}XX}$) stands for contributions from the 1PI diagram (the counter term). 
The fuzzy Higgs boson $h_{\rm eff}$ can be  regarded as  $h$ in the flavor basis. 
Hence, the tree level coupling $\Gamma^{\rm tree}_{h {\rm eff}XX}$ coincides with that of the SM,
i.e., $\Gamma^{\rm tree}_{h {\rm eff}XX}=g_{hXX}^{\rm SM}$, 
where $g_{hXX}^{\rm SM}$ denotes the tree level coupling in the SM, e.g., $g_{hbb}^{\rm SM}=m_{b}/v$ and $g_{hWW}^{\rm SM}=m_{W}/v$. 
The deviations from the SM prediction arise by higher-order corrections.
We note that, in addition to the loop corrections to $\Gamma^{\rm tree}_{h {\rm eff}XX}$, the correction to the propagator in the Higgs production $e^+ e^- \to Zh_{\rm eff} $, $\delta$ or the effective mixing $\a_{\rm eff}$ should be also incorporated in the deviation of the Higgs couplings. It can be written by $\cos[\a_{\rm eff}] g_{hXX}^{\rm SM}$. This effect will be not included in this part, and we will see if the other loop contributions can introduce a sizable deviation. 

The counter-term $\delta \Gamma^{}_{h {\rm eff}XX}$ can be derived by shifting parameters and fields in  the interaction term $g_{hXX}^{\rm SM}hXX$. 
We perform this in the mass basis and obtain the counter terms for the $h_{\rm eff}XX$  vertex by returning to the flavor basis with the corresponding orthogonal transformation. 
The shift, {or kinetic normalization}, of the Higgs boson fields  are defined by
\begin{align}
\begin{pmatrix}
\phi_{1} \\ \phi_{2} 
\end{pmatrix}
& \to 
{\hat{Z}^{1/2}}
\begin{pmatrix}
\phi_{1} \\ \phi_{2} 
\end{pmatrix}
\\ 
\text{with} ~\hat{Z}^{1/2} &\equiv \begin{pmatrix}
1+\frac{1}{2}\delta Z_{{11}} & \frac{1}{2}\delta Z_{{1}{2} } \\
 \frac{1}{2}\delta Z_{{2}{1} }& 1+\frac{1}{2}\delta Z_{{22}} \\
\end{pmatrix}. 
\end{align}
where $\delta Z_{ij}$ ($i,j=1,2$) are the wave functions renormalization constants (WFRCs) for the Higgs bosons $\phi_1$, $\phi_2$. 
Applying the shift of $\phi_1,\phi_2$,  a SM field $X$ and the coupling $g_{hXX}^{\rm SM}$ yield the counter term 
\begin{align}
\delta \Gamma_{h_{\rm eff}XX}&=
g_{hXX}^{\rm SM}\Bigg\{\cos^{2}\alpha\frac{\delta {Z_{11}}}{2}+\sin^{2}\alpha\frac{{\delta Z_{22}}}{2} \notag \\
&-\sin\alpha\cos\alpha\left(\frac{\delta Z_{12}}{2}+\frac{\delta Z_{21}}{2}\right)\Bigg\}+\delta \Gamma_{h_{\rm eff}XX}^{\rm rem},
\end{align}
where $\delta \Gamma_{h_{\rm eff}XX}^{\rm rem}$ denotes the counter term contributions except for the Higgs bosons, {which are the same as the SM at the 1-loop level.} {Here the first term comes from $\(R \hat{Z}^{1/2} R^{-1}-\mathbf{1}\)_{hh} g^{\rm SM}_{hXX}$.}
For completeness, we show the explicit formulae of $\delta \Gamma_{h{\rm eff}XX}^{\rm rem}$ for the $hbb$ coupling and the $hWW$ coupling, 
\begin{align}
\delta \Gamma_{h{\rm eff}bb}^{\rm rem}&=-\frac{m_{f}}{v}
\left(\frac{\delta m_{f}}{m_{f}}-\frac{\delta v}{v}+\frac{\delta Z_{b,L}}{2}+\frac{\delta Z_{b,R}}{2}\right),  \\
\delta \Gamma_{h{\rm eff}WW}^{\rm rem}&=\frac{2m_{W}^{2}}{v}
\left(\frac{\delta m^{2}_{W}}{m^{2}_{W}}-\frac{\delta v}{v}+\delta Z_{W} \right). 
\end{align}
For the definition of counter terms of EW VEV, the masses,  the WFRCs for $b$ and $W$, we refer, e.g.,  \cite{Denner:1991kt}. 
 Remarkably, the counter term from the mixing angle $\alpha$ is canceled out in those expressions. 
 
In actual numerical calculations for the deviations in the Higgs boson couplings, we omit $\delta \Gamma_{h_{\rm eff}XX}^{\rm rem}$ as well as the 1PI diagram contributions  $\Gamma^{\rm 1PI}_{h {\rm eff}XX}$ {since the $s$ contribution does not appear in the 1-loop level.\footnote{The exception is the Higgs boson self-coupling $h_{\rm eff}h_{\rm eff}h_{\rm eff}$.  For this coupling, 1P1 diagrams involve dominant contributions {by $s$}. This effect is also suppressed by the small $|\lambda_P|$.} 
{In Refs.~\cite{Kanemura:2004mg,Kanemura:2014dja,Kanemura:2015mxa} and Refs.~\cite{Kanemura:2015fra,Kanemura:2016lkz} one can find the calculations for the 1-loop corrections to the Higgs boson couplings in two Higgs doublet model and Higgs singlet model, respectively.} 
} 

The WFRCs $\delta {Z}_{ij}$ are determined by the on-shell conditions for the Higgs bosons in the mass basis $\phi_1$ and $\phi_2$~\cite{Krause:2016oke,Kanemura:2017wtm}. 
{
They are written by
\begin{align}
\delta Z_{11}&=-\Pi_{11}'(m_{1}^{2})\;,\quad
\delta Z_{22}=-\Pi_{22}'(m_{2}^{2})\;, \\
\delta Z_{12} &=-2\frac{\tilde{\Pi}_{12}(m_2^2)}{m_2^2-m_1^2}\;,\quad
\delta Z_{21} =2\frac{\tilde{\Pi}_{12}(m_1^2)}{m_2^2-m_1^2}\;,
\end{align}
where $\Pi_{ii}'(m_i^2)\equiv d\Pi_{ii}(Q^2)/dQ^{2}|_{Q^2=m_i^2}$ and $\tilde{\Pi}_{12}$ is defined by
 \begin{align}
 \tilde{\Pi}_{12}={\Pi}_{12}-2\sin\alpha\cos\alpha\left(\frac{\delta T_{h}}{v}-\frac{\delta T_{s}}{v_{s}}\right)
 \end{align}
 with $\delta T_{h}=\cos\alpha \delta T_{1}-\sin\alpha \delta T_{2}$ and $\delta T_{s}=\sin\alpha\delta T_{1}+\cos\alpha\delta T_{2}$.
}
For the renormalization of tadpoles, the standard tadpole scheme is used, where the counter terms for the tadpole are set in the way that the renormalized tadpoles disappear at the 1-loop level, i.e., $\hat{T}_{i}=T_{i}+\delta T_{i}=0$. 

{We define the deviations of the Higgs boson couplings by}
\begin{align}
{
\kappa_X=\frac{\Gamma_{h{\rm eff} XX}}{\Gamma^{\rm SM}_{h XX}}, }
\end{align}
{where the $\Gamma^{\rm SM}_{h XX}$ denotes the renormalized vertex function in the SM. }
With the obtained concrete expressions for $\delta {Z}_{ij}$ , the deviations in the Higgs boson couplings can be written as 
\begin{align}\label{eq:dev}
\kappa_{X}&\sim \Big(1-\frac{\cos^{2}\alpha}{2} \left.\Pi_{11}'(m_{1}^{2})\right|_{\rm fin.}-\frac{\sin^{2}\alpha}{2} \left.\Pi_{22}'(m_{2}^{2})\right|_{\rm fin.} \notag \\
&-\frac{\sin\alpha\cos\alpha}{m_{2}^{2}-m_{1}^{2}}
\left\{\left.\tilde\Pi_{12}(m_{1}^{2})\right|_{\rm fin.}-\left.\tilde{\Pi}_{12}(m_{2}^{2})\right|_{\rm fin.} \right\}
\Big) \notag \\
&/\left(1-\frac{1}{2}\left.\Pi^{{\rm SM}\prime}_{hh}(m_{h}^{2})\right|_{\rm fin.}\right),
\end{align}
{where $\Pi^{{\rm SM}}_{hh}$ denotes the self energy for the Higgs boson $h$ in the SM.}
We take finite part of each self energy in Eq.~\eqref{eq:dev}
{, which is denoted by $\Pi^{(\prime)}_{ij}(Q^2)|_{\rm fin.}$.}
UV divergence cancels out if one involves the remaining contributions $\Gamma^{\rm 1PI}_{h{\rm eff}XX}+ \delta \Gamma^{}_{h{\rm eff}XX}$.

From numerical calculation {with {\tt LoopTools}~\cite{Hahn:1998yk}, we show deviation,  $\k_X-1$, in Fig.~\ref{fig:2} by varying $v_s$ and $\epsilon$ in the upper and lower panels, respectively. We fix $\epsilon=10\GEV^2,100\GEV^2,1000\GEV^2$ from top to bottom in the upper panel and $v_s=10\GEV,100\GEV,1000\GEV$ from bottom to top in the bottom panel.
{The small contribution to the deviation of the Higgs coupling in the regime is due to the small portal coupling $\l_P$ that leads to a small mass difference. This interpretation is consistent with the dependence of $|\epsilon|$. Also one can see that $|\k_X-1|$ is not proportional to $v_s^2$ i.e. it is not suppressed by $\G_s.$ Thus the suppression of the deviation has a different origin from the suppression of the mixing effect in the main part. The deviations of the couplings are negligible in the regime of our interest since $|\epsilon|$ is usually not very large in the fuzzy Higgs boson scenario (at least when the dark sector is perturbative).
}

\begin{figure}[!t]
\begin{center}  
   \includegraphics[width=.45\textwidth]{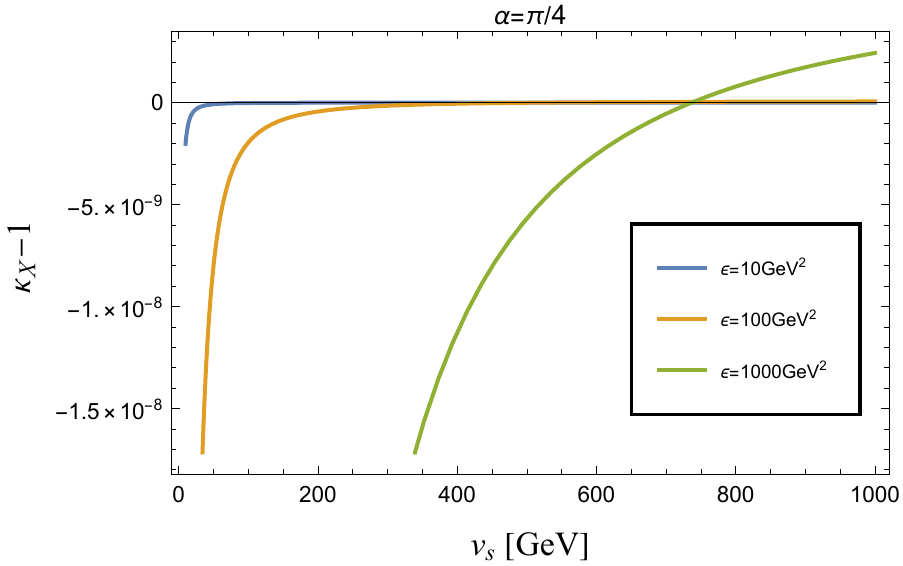}
   \includegraphics[width=.45\textwidth]{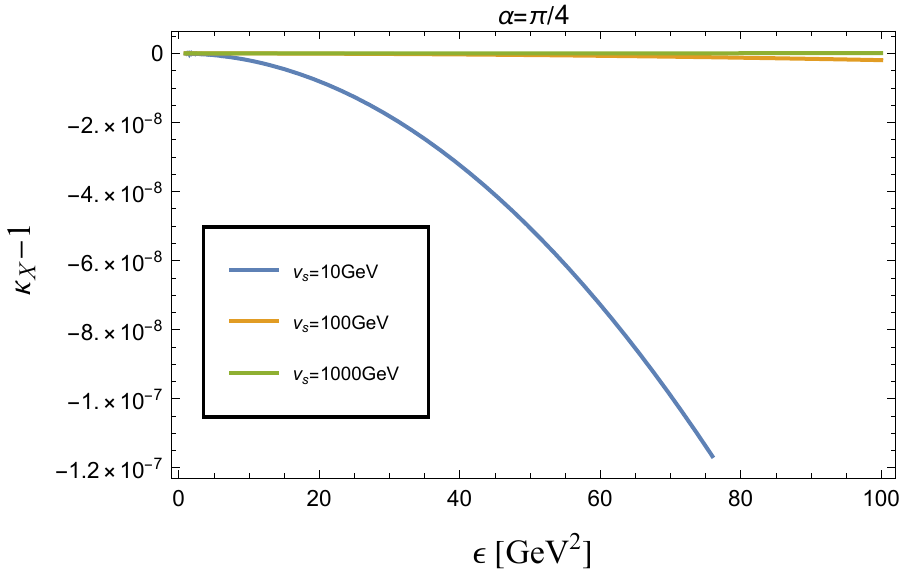}
      \end{center}
\caption{The deviation of the Higgs coupling, $\k_X-1$ by varying $v_s$ (upper panel) and $\epsilon$ (lower panel) with fixing $\alpha=\pi/4$.  } \label{fig:2} 
\end{figure}

\bibliography{Zeno}
\end{document}